# On the Phase Tracking Reference Signal (PT-RS) Design for 5G New Radio (NR)

Yinan Qi[1], Mythri Hunukumbure[1], Hyungju Nam[2], Hyunil Yoo[2], Saidhiraj Amuru[3]
[1]Samsung Electronics R&D Institute UK, Staines, Middlesex TW18 4QE, UK
[2] Samsung Electronics, Suwon, Korea
[3] Samsung R&D Institute, Bengaluru, India
yinan.qi, mythri.h, hyungju.nam, hyunil.yoo and a.saidhiraj@samsung.com

*Abstract*—The volume of mobile data traffic has been driven to an unprecedented high level due to the proliferation of smartphones/mobile devices that support a wide range of broadband applications and services, requiring a next generation mobile communication system, i.e., the fifth generation (5G). Millimeter wave (mmWave) bands can offer much larger available spectrum bandwidth and thus are considered as one of the most promising approaches to significantly boost the capacity in 5G NR. However, devices and network radio nodes operating on mmWave bands suffer from phase noise and without correction of phase noise, the performance of the network could potentially suffer significant losses. In this paper, we investigate the effects of phase noise and provide comprehensive solutions to track the phase noise by using phase tracking reference signals (PT-RS), as currently standardized in 3GPP Release 15. The design aspects such as PT-RS pattern, interference randomization, multi-TRP operation, etc., are investigated and evaluation results are also provided.

## I. INTRODUCTION

5G network is expected to offer system access and services that have different characteristics and connectivity control for future services. In this regard, it needs to be highly flexible and tailored towards the new requirements. The foundation of this next generation cellular network is 5G New Radio (NR) [1]-[2], a global 5G standard for a new OFDM-based air interface designed to support the wide variation of 5G device-types, services, deployments and spectrum. The most apparent transformation taking place with 5G NR is the move towards higher millimeter wave (mmWave) frequencies as a very promising approach to significantly boost the capacity of 5G. However, mmWave devices and network access points suffer from severe phase noise mainly due to the mismatch of transmitter and receiver frequency oscillators [3]-[4]. Basically, phase noise is caused by noise in the active components and lossy elements which is up-converted to the carrier frequency. Frequency synthesizers generally consist of a reference oscillator (or clock), a voltage controlled oscillator (VCO), and a phase-locked loop (PLL) with frequency divider, phase-frequency detector charge pump, and loop filter. In this regard, phase tracking reference signal (PT-RS) is introduced in 5G NR to tracking the phase and mitigate the performance loss due to phase noise [5]-[6].

The European Commission funded, 5GPPP phase I mmMAGIC project [7]-[8] investigated the phase noise modelling at mmWave frequencies and the initial design solutions of PT-RS insertion (involving time and frequency densities) [9]. The current phase II ONE5G project [10]-[11] takes a more holistic view on 5G RAN design, covering both cmWave (below 6 GHz) and mmWave spectrum components. In ONE5G the reference signal design is investigated against the challenges posed by emerging technologies like massive MIMO [11].

In this paper, we provide a comprehensive solution covering various aspects of PT-RS design issues, e.g., PT-RS pattern, interference randomization, etc. Some of the investigated issues have been captured in 5G NR standard [5]-[6] and some issues are potential enhancements for further development of 5G. In section II, phase noise models are introduced and OFDM based signal model considering phase noise is studied in section III. Various PT-RS design issues are investigated in section IV and evaluation results are provided as well. The paper is concluded in section V.

## II. PHASE NOISE MODELS

The characteristic of phase noise is usually explained from its power spectrum. Thus, several ways to make good approximation to practical phase noise spectra are developed for analysis. The simplest one is a single pole/zero model which is adopted in IEEE P802.15 [12]. However, it is a simple linear model for PLL so that it does not consider other phase noise sources. In [13], a new model considering three main noise sources such that reference clock, PLL, and VCO is proposed but loop bandwidth cannot be tuned easily. Therefore, as a compromised solution between easiness of analysis and good approximation to reflect practical phase noise characteristic, we proposed the multi-pole/zero model which is extended from a single pole/zero model by adding more pole/zero frequency terms as follows:

$$S(f) = \text{PSD0} \prod_{n=1}^{N} \frac{1 + \left(\frac{f}{f_{z,n}}\right)^2}{1 + \left(\frac{f}{f_{p,n}}\right)^2}, \quad (1)$$

where PSD0 is the power spectral density for zero frequency ($f$=0) in dBc/Hz, $f_{z,n}$ are zero frequencies, and $f_{p,n}$ are pole frequencies. The multi-pole/zero model has some advantages as follows:

• Provides an easy framework to convert the PSD of analog phase noise to that of discrete-time phase noise (i.e., baseband version) for simulation by using the bilinear transform with given pole/zeros.

• Practical phase noise power spectra can be well approximated with a few pole/zeros

Table 1 shows two parameter sets which are obtained from practical oscillators operating at 30GHz and 60GHz, respectively [14].

**Table 1**

|  | Parameter Set-A | Parameter Set-B |
|---|---|---|
| Carrier frequency ($f_{c,base}$) | 30GHz | 60GHz |
| PSD0 (dBc/Hz) | -79.4 | -70 |
| Fp (MHz) | [0.1, 0.2, 8] | [0.005, 0.4, 0.6] |
| Fz (MHz) | [1.8, 2.2, 40] | [0.02, 6, 10] |

In addition, if the operating carrier frequency is changed, the PSD is shifted by $20\log_{10}(f_c/f_{c,base})$ dBc/Hz. Figure 1 shows the PSDs of the two parameter sets in 4GHz, 30GHz, and 70GHz center frequencies, respectively.

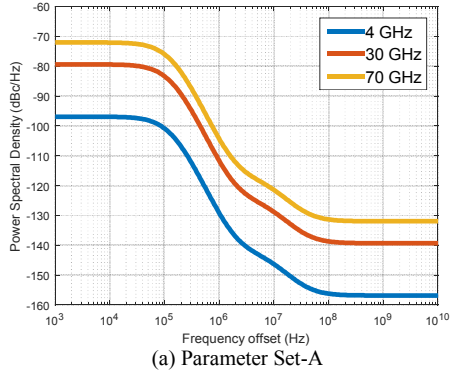
(a) Parameter Set-A

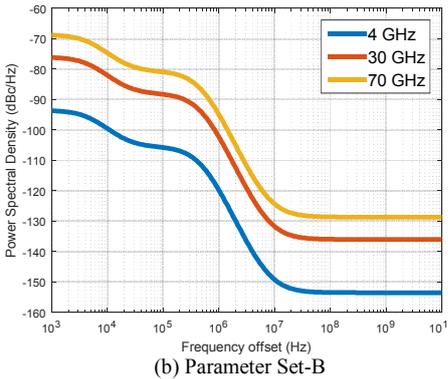
(b) Parameter Set-B

Figure 1. Phase noise power spectral density

### III. OFDM BASED SIGNAL MODEL

When the mismatch of oscillator frequencies between transmitter and receiver occurs, the frequency difference implies a shift of the received signal spectrum at the baseband. In OFDM, this creates a misalignment between the bins of FFT and the peaks of the sinc pulses of the received signal. This breaks orthogonality between the subcarriers so that results in a spectral leakage between them. Each subcarrier interferes with every other (although the effect is dominant between adjacent subcarriers), and as there are many subcarriers this is a random process equivalent to Gaussian noise. Thus, this frequency offset lowers the SINR of the receiver. An OFDM receiver will need to track and compensate phase noise.

The baseband received signal in the presence of only phase noise, assumed that there is no additive white Gaussian noise (AWGN), is given as the following equation:

$$y[n] = \frac{1}{\sqrt{N}} \sum_{k=0}^{N-1} S_k e^{j\frac{2\pi}{N}kn} e^{j\theta[n]}, \quad (2)$$

where the transmitted signal is multiplied by a noisy carrier $\exp(j\theta[n])$.

The received signal is passed through the FFT in order to obtain the symbol transmitted on the m-th subcarrier in the OFDM symbol as follows:

$$\hat{S}_m = \frac{1}{N} \sum_{n=0}^{N-1} y[n] e^{-j\frac{2\pi}{N}mn}$$

$$= S_m \frac{1}{N} \sum_{n=0}^{N-1} e^{j\theta[n]} + \frac{1}{N} \sum_{k=0, k\neq m}^{N-1} S_k \sum_{n=0}^{N-1} e^{j\frac{2\pi n}{N}(k-m)} e^{j\theta[n]} \quad (3)$$

Since the first term of the right hand side in (3) (i.e., mean of $\exp(j\theta[n])$ during one OFDM symbol duration) does not depend on subcarrier index m, it is called common phase error (CPE). This term causes common phase rotation in constellations of received symbols. The CPE can be estimated from the reference signals and removed. And the second term causes inter-carrier interference (ICI). The ICI due to phase noise creates a fuzzy constellation as shown in Figure 2.

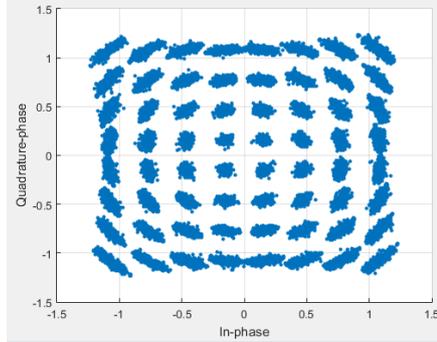
Figure 2. Phase noise impacts on constellations (64QAM)

EVM (Error Vector Magnitude) is defined as square root of the ratio of the mean error vector power to the mean reference power expressed in percent as below

$$EVM(dB) = 10\log_{10}\left(\frac{P_{error}}{P_{reference}}\right) \quad (4)$$

where $P_{error}$ is the root mean square (RMS) amplitude of the error vector and $P_{reference}$ is the RMS amplitude of the reference vector. Tx EVM requirement for 5G NR is given as follows.

**Table -2**

| Modulation scheme | Required EVM [%] | Required EVM [dB] |
|---|---|---|
| QPSK | 17.5 % | -15.14 dB |
| 16QAM | 12.5 % | -18.06 dB |
| 64QAM | 8 % | -21.93 dB |

It has been shown in [14] that with CPE compensation only the required EVM can be achieved at least for the lower end of mmWave, e.g., 30GHz. Therefore, phase tracking reference signal (PT-RS) is introduced mainly to compensate CPE, which is the focus of this paper. PT-RS can also be used for ICI mitigation in higher frequency bands and potentially for CFO and Doppler estimation but it is out of the scope of this paper.

## IV. PT-RS DESIGN

In this section, we investigate the PT-RS design aspects, such as pattern design, interference randomization, enhancements for CoMP operation, etc. Evaluation results are also presented and the parameters are defined in [15].

### A. PT-RS pattern

There is a trade-off between phase tracking accuracy and signaling overhead. If the density of PT-RS is high, phase tracking accuracy is high and CPE can be better compensated to achieve better performance. However, higher PT-RS density also means larger signaling overhead, which might lead to lower spectrum efficiency or effective transmission rate, i.e., number of information bits transmitted per second per Hz. Figure 3 shows two different PT-RS time densities.

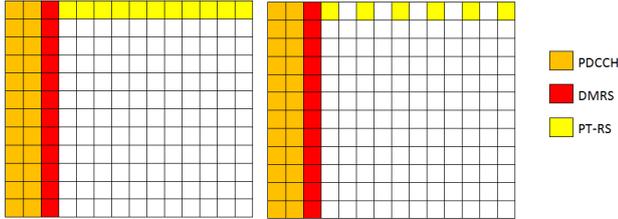

Fig. 3 PT-RS time/frequency density

In this section, the performance in terms of BLER for different PT-RS density in the time domain is presented and analysed. BLER performance is illustrated in Figure 4 and 5 for the cases where the number of allocated physical resource blocks (PRBs) is 32.

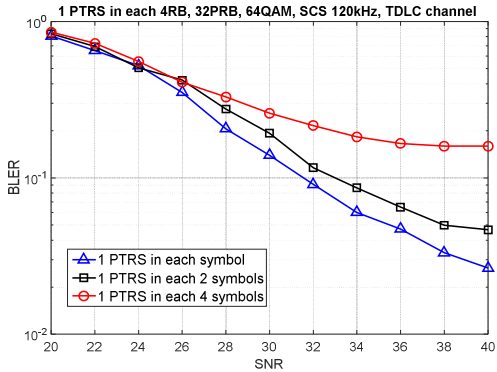

Fig. 4 64QAM

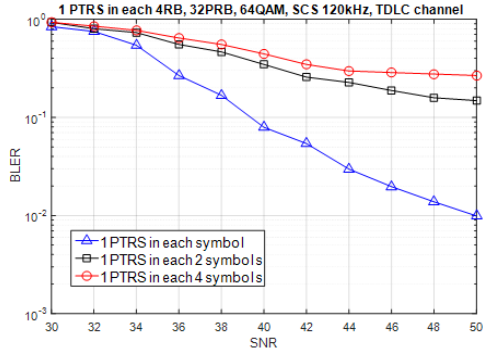

Fig. 5 256 QAM

Figure 4 and 5 show that reduction of PT-RS density in time domain will degrade the BLER performance regardless of modulation order. However, the performance degradation is particularly significant when time density is reduced from 1 to 2, i.e., from PT-RS for each OFDM symbol to PT-RS for every other OFDM symbol in the case of 256 QAM. In such a case, even though PT-RS signaling overhead is halved for time density 2, i.e., more information data bits can be transmitted in each RB, the effective information data transmission rate suffers performance loss due to much worse BLER. On the contrary, for 64QAM the degradation due to time density reduction is much less. Considering the PT-RS signaling overhead is reduced by half from time density 1 to 2, the spectrum efficiency might actually be improved. Therefore, the time density of PT-RS can be a function of modulation order and it should increase with higher modulation order.

According to the allocated RBs, we also investigate the BLER performance with various PT-RS density in the frequency domain. Figure 6 and 7 shows the BLER performances of which the allocated PRBs are 32PRB and 8PRB, respectively. In order to focus on the performance according to frequency-domain density, it is assumed that PT-RS is allocated in every OFDM symbol in time domain.

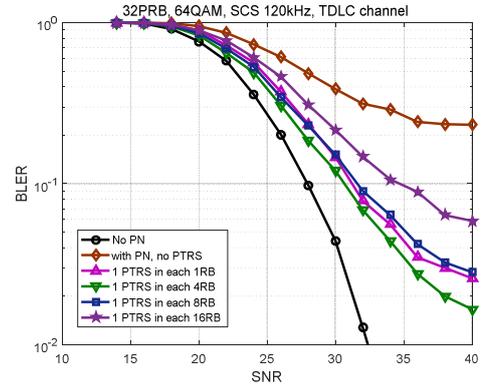

Fig.6 32 PRB

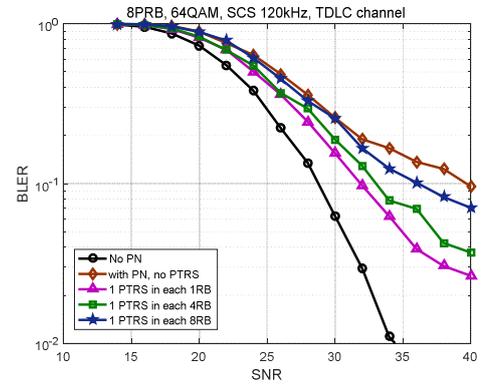

Fig. 7 PRB

In cases of 32PRB, the BLER performance of the 1 PT-RS in each 4RB shows the better performance compared to the 1 PT-RS in each 1RB. These results represent that the excessive PT-RS density just decrease the effective code rate for data transmission. In the 8 PRB case, the BLER performance of the 1 PT-RS in each 1RB has the best performance. We can infer that the phase noise compensation is more important when the small PRB is allocated. In addition, higher PT-RS density also means larger signaling overhead. Considering the trade-off between the PT-RS overhead and performance, the proper number of subcarriers can be different to accurately estimate CPE according to allocated number of RBs from the results. From the efficient resource management perspective, it is beneficial to configure PT-RS pattern in frequency domain based on scheduled bandwidth and the density should decrease with larger number of allocated bandwidth.

### B. Interference randomization

Since only phase different needs to be tracked using PT-RS, the receiver does not need to know the amplitude of PT-RS. Therefore, unlike other reference signals, e.g., DMRS, where the reference signals are formed by a pseudo-random sequence of symbols, PT-RS reference signals can use the exact same symbol. In MU-MIMO, PT-RS can be configured to each user and it is possible that the same subcarrier is used for multiple users as shown in Fig. and therefore PT-RS collision happens, which will degrade CPE compensation performance for two reasons: 1) the interference pattern is not completely random since the same symbol is used for PT-RS; 2) the interference level is higher when the power of PT-RS is boosted for more accurate CPE compensation. In such a case, it would be better to avoid PT-RS collision so that the interference is from data symbols from another user and thus the interference is randomized without power boosting. This can be avoided by introducing an RB level offset when configuring PT-RS for each user.

As aforementioned, the frequency density of PT-RS can be every four RBs. Assuming the number of RBs allocated to a user is $N_{RB}$ and the frequency density is $d_f$, the exact RB location of PTRS still cannot be identified as shown in Fig, where $P_{offset}$ is the PTRS RB offset.

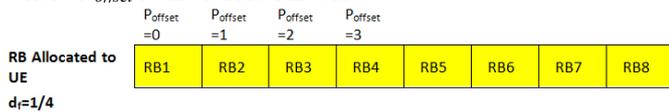

Figure 8 RB offset

In the above example, there are 4 possible values to choose $P_{offset}$. Without knowing the exact offset value, UE cannot locate the PT-RS correctly. To achieve interference randomization, e.g., avoiding PT-RS interfering with PT-RS, different PT-RS offset values should be configured for different users, i.e., user-specific configuration based offset should be employed. Evaluation results for interference randomization in this regard are shown in Fig. 9 where MCS level is 25 with modulation order 64 QAM. The interference level from the interfering UE is assumed to be 40dB. Almost 2 dB gain can be observed.

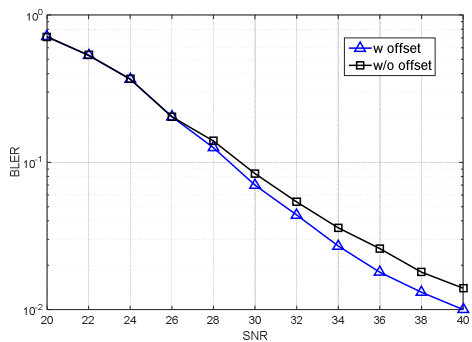

Fig. 9 Interference randomization.

### C. PTRS Insertion for DFT-s-OFDM waveform

In NR, as opposed to LTE, both DFT-s-OFDM and CP-OFDM waveforms are supported for the uplink transmissions. These PTRS signals are necessary for both these waveforms. The PTRS insertion follows a common framework for both the downlink and uplink in case of CP-OFDM waveforms. IN the case of DFT-s-OFDM waveform, 3GPP considered two types of insertion mechanisms for these RS, namely pre-DFT insertion and post-DFT insertion. In the first mechanism, PTRS signals are inserted in the frequency domain before DFT pre-coding so that the resulting waveform still maintains a single carrier property. In the latter mechanism, the RS are inserted after DFT pre-coding of the data symbols via various mechanisms, such as puncturing for example. The PAPR of such a mechanism can however be controlled by using some simple signal processing techniques (the details of which are omitted for brevity) [16], [17]. A representative figure is shown in Fig. 10, wherein X and K are two parameters that indicate the number of chunks of PTRS samples (in a pre-DFT insertion mechanism) which are equally spaced and the number of samples within each chunk respectively.

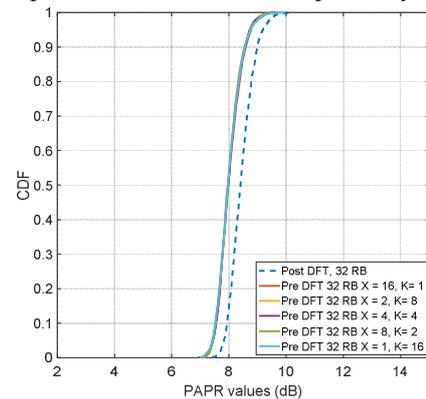

Fig 10. PAPR performance comparison of PT-RS insertion schemes for 32 RB allocation in a DFT-s-OFDM waveform

While pre-DFT indeed has lower PAPR, the post-DFT insertion mechanism can help to achieve a similar PTRS insertion mechanism as CP-OFDM waveform since the RS is inserted after DFT pre-coding and before IFFT operation. Also as shown in [16] and [17], post-DFT PTRS based mechanisms can achieve better performance in terms of the block-error rate compared to pre-DFT mechanisms. However, these performance results differed across various companies' evaluation results due to various assumptions. Hence, after several discussions, 3GPP agreed to support pre-DFT insertion mechanism for PTRS in the DFT-s-OFDM waveform for Rel-15. The pattern and density of the insertion of PTRS in pre-DFT insertion is defined as shown in the table below where $N_{RB}$ is the number of scheduled resource blocks for the UE, and the values $N_{RBi,\ i=0,1,2,3,4}$ are configured by higher layers.

Table 3

| Scheduled bandwidth | Number of PT-RS chunks (X) | Number of samples per PT-RS chunk (K) |
|---|---|---|
| $N_{RB0} \leq N_{RB} < N_{RB1}$ | 2 | 2 |
| $N_{RB1} \leq N_{RB} < N_{RB2}$ | 2 | 4 |
| $N_{RB2} \leq N_{RB} < N_{RB3}$ | 4 | 2 |
| $N_{RB3} \leq N_{RB} < N_{RB4}$ | 4 | 4 |
| $N_{RB4} \leq N_{RB}$ | 8 | 4 |

In time domain, these PTRS locations can be configured to be either present in every symbol or every other symbol.

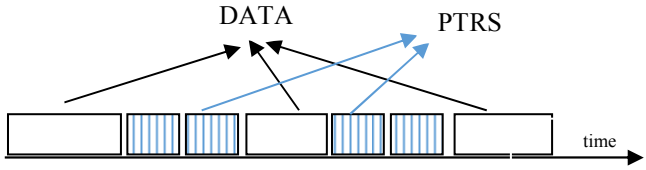

Fig. 11 An example representation for the PTRS locations (shaded) before DFT pre-coding with X=2, K=2.

*D. CoMP and multi-TRP Operation*

In CoMP, a single UE can be supported by multiple neighbouring gNBs, thus turning the interference in a single supporting gNB case to useful signals. Also in 5G NR, large numbers of antenna elements are expected to be used at the gNB. These antenna elements are usually grouped as panels, where the carrier signals feeding into antenna panels are driven by separate oscillators. This is termed as multi-TRP (Transmit Panel) operations. Thus in a phase noise sense, the UE in both the CoMP and multi-TRP scenarios receive multiple carrier signals driven by different oscillators, which need individual compensation.

In designing PT-RS for multi-TRP or CoMP transmissions, the orthogonal allocation of PT-RS carrying resource elements (REs) is the most robust option. An example with 2 CoMP gNBs or multi-panels is shown below in Figure 12.

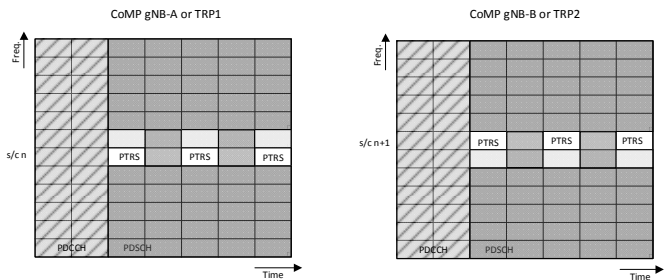

Fig. 12 PT-RS allocation in CoMP or Multi-TRP operations

The orthogonality for the PT-RS can be provided in the frequency domain, as shown in the example. The increase in signaling overhead is a valid concern when there is a higher number of gNBs in the CoMP set or many transmit panels. A possible implementation solution for the multi-TRP case is discussed below, in a multi-user (MU) MIMO context.

As discussed above, the typical CPE caused by the phase noise rotates the constellations by a limited margin, so only the higher order modulation schemes are impacted by the CPE. The users with higher MCS receive good SNR levels and are usually located closer to the gNB. When the MU-MIMO user sets are grouped, there will be higher and lower MCS users in these groups. If the PT-RS is transmitted without power boosting and with a wider beam than the narrow beam data transmissions, the received EIRP (effective isotropic radiated power) for the PT-RS will be lower than for data transmissions. The lower MCS users, who are generally further away from the cell centre, will receive PT-RS with a much lower effective power and will be able to discard PTRS (they will not need CPE correction) as interference. They will be able to request the gNB to allocate data within these REs, transmitted through narrow beams. The same REs are used for PT-RS in the wider beam transmissions for the benefit of higher MCS users, for whom the same REs will be kept vacant in the narrow beam data transmissions. With this effective power discrimination Non-orthogonal multiplexing of PT-RS and data is possible for the MU-MIMO configurations, which effectively increases the system spectral efficiencies.

## V. CONCLUSIONS

A comprehensive analysis on the phase noise modelling and compensation with PT-RS as currently standardized in 3GPP NR is presented in this paper. The impact of different PT-RS densities in the time and frequency domains for different MCS schemes and allocation bandwidths (PRB sizes) are discussed, supported by simulation results. The effects of interference randomization through providing different PT-RS offsets are also shown through simulations. The possible increase in the signaling overhead when multi-panel TRP or CoMP systems are allocated with orthogonally multiplexed PT-RS can be a major factor in performance limitation. A possible implementation solution in the MU-MIMO set-up is presented for this issue.

Currently in 3GPP-NR Release 15, the specifications related to PT-RS are being finalized, mainly to support eMBB use cases through CP-OFDM. With Release 16 intended to provide more support for new services such as integrated access and backhaul, non-terrestrial network, etc., the research on Phase Noise will move to new grounds. We would continue our PN related research to cover these new aspects.

## ACKNOWLEDGEMENTS

The European Commission funding is acknowledged for the phase I mmMAGIC project (grant agreement 671650) and the phase II ONE5G project (ICT 760809), through which part of the research leading to this paper was conducted.